\documentclass[12pt,preprint]{aastex}

\pdfoutput=1

\usepackage{float}
\usepackage{natbib}
\usepackage{enumerate}
\usepackage{multirow}
\usepackage{array}

\def\plotone#1{\centering \leavevmode
\includegraphics[clip=, width=.85\columnwidth]{#1}}

\newcommand{\cN}[1]{\mathcal{N}}

\def\gsim{\;\rlap{\lower 2.5pt
 \hbox{$\sim$}}\raise 1.5pt\hbox{$>$}\;}
\def\lsim{\;\rlap{\lower 2.5pt
   \hbox{$\sim$}}\raise 1.5pt\hbox{$<$}\;}

\begin{document}


\title{%
Binary Evolution Leads to Two Populations of White Dwarf Companions
}

 \author{
David S. Spiegel,
}

 \vspace{0.5\baselineskip}
 
 \email{
dave@ias.edu
}

\begin{abstract}
Planets and other low-mass binary companions to stars face a variety
of potential fates as their host stars move off the main sequence and
grow to subgiants and giants.  Stellar mass loss tends to make orbits
expand, and tidal torques tend to make orbits shrink, sometimes to the
point that a companion is directly engulfed by its primary.
Furthermore, once engulfed, the ensuing common envelope (CE) phase can
result in the companion becoming fully incorporated in the primary's
envelope; or, if the companion is massive enough, it can transfer
enough energy to eject the envelope and remain parked in a tight orbit
around the white dwarf core.  Therefore, ordinary binary evolution
ought to lead to two predominant populations of planets around white
dwarfs: those that have been through a CE phase and are in
short-period orbits, and those that have entirely avoided the CE and
are in long-period orbits.
\end{abstract}

\section{Introduction}
\label{sec:intro}

Many intermediate-mass stars have either stellar companions or brown
dwarf or planetary companions \citep{duquennoy+mayor1991,
  raghavan_et_al2010, schneider_et_al2011, wright_et_al2011}.  These
systems can sometimes remain dynamically quiet for hundreds of
millions or billions of years, and then undergo relatively rapid
changes when the primary star evolves off the main sequence.

Various recent works have investigated the post-main-sequence fates of
two-body systems \citep{villaver+livio2007, carlberg_et_al2009,
  villaver+livio2009, nordhaus_et_al2010} and of many-body systems
\citep{veras_et_al2011, veras_et_al2012, kratter+perets2012}.  The
present work is generally concerned with the former type of system ---
i.e., the evolution of a single star and a single low-mass companion.

The status of planetary systems around post-main-sequence stars is
starting to come into focus.  \citet{zuckerman_et_al2010} find
evidence of heavy-element atmospheric pollution in $\sim$1/3 of DB
white dwarfs, which they interpret as due to accretion of tidally
shredded asteroids that were presumably scattered onto
high-eccentricity orbits by distant planets.  \citet{maxted_et_al2006}
found evidence of perhaps the best-characterized substellar companion
to a white dwarf, via high-resolution spectroscopy that revealed
radial velocity variations with period 2~hours with semiamplitudes of
$\sim$28~km~s$^{-1}$ in the primary and $\sim$188~km~s~$^{-1}$ in the
secondary.  They infer a secondary mass of $\sim$50~$M_J$ (where $M_J$
is the mass of Jupiter) --- clearly above the canonical ``planet''
mass but less massive than the hydrogen-burning mass limit
\citep{burrows_et_al2001}.  Provocative evidence of very close
low-mass companions to an evolved star was reported by
\citet{charpinet_et_al2011}, who claim to find two sub-Earth-radius
planets around the subdwarf star KIC 05807616 that are both orbiting
inside 1.7~$R_\sun$ (where $R_\sun$ is the radius of the Sun).  If
these planet candidates are actually planets, this discovery shows
that exotic and theoretically unanticipated post-main-sequence
planetary systems are possible.  \citet{hogan_et_al2009} report a
nondetection of warm companions around 23 nearby white dwarfs and
suggest that $\lesssim$5\% of white dwarfs have companions with
effective temperatures greater than 500~K between 60~AU and 200~AU in
projected separation.  Several years ago, the pulsating white dwarf
GD-66 was found to exhibit timing variations in pulsations that seemed
to be consistent with light-travel-time delays caused by orbital
motion around the common center of mass with a companion on a
$\sim$4.5-year orbit \citep{mullally_et_al2007, mullally_et_al2008,
  mullally_et_al2009}.  However, recent high-precision astrometric
observations of this system, an analysis of {\it Spitzer} observations
of the system, and follow-up observations by the team of original
discovery, have complicated the planetary hypothesis for this system
(\citealt{farihi_et_al2012}; J. Hermes 2012, private communication).
Finally, \citet{johnson_et_al2011} has found a number of giant planets
around slightly evolved, subgiant stars (see also
\citealt{sato_et_al2008, bowler_et_al2010}), some of which might be
near the edge of engulfment or repulsion.  A number of lines of
evidence, then, point to the existence of planets around evolved
stars; even if some of the candidate systems eventually turn out not
actually to be planets, it still seems clear that several tens of
percent of white dwarfs have planets around them
\citep{zuckerman_et_al2010}.

It is worthwhile to recognize that the existence of planets orbiting
post-main-sequence stars should be no surprise, given that that the
very first planets discovered beyond our solar system were found
around the millisecond pulsar PSR1257+12 \citep{wolszczan+frail1992}.
Whether planets around stellar remnants formed after the death of the
primary, or were present during the main-sequence phase and survived
stellar evolution, remains an open question.  It seems likely that
both processes occur \citep{hansen_et_al2009, tutukov+fedorova2012};
although there might be more reason to believe that planets could form
around a pulsar than around a white dwarf, there is no consensus as
yet.

Many jovian companions to main-sequence stars will become highly
irradiated as their primaries evolve off the main sequence, thereby
turning them into ``hot Jupiters,'' of sorts --- or red-giant hot
Jupiters, as described by \citet{spiegel+madhusudhan2012}.  The
atmospheres of these companions might become transiently polluted by
accretion both of the evolved star's wind and of dust and
planetesimals \citep{spiegel+madhusudhan2012, dong_et_al2010}.  Some
of these planetary companions, or somewhat higher mass companions,
will eventually be swallowed by their stars, which might contribute to
the formation and morphology of planetary nebulae
\citep{nordhaus+blackman2006, nordhaus_et_al2007} and to the formation
of highly magnetic white dwarfs \citep{tout_et_al2008,
  nordhaus_et_al2011}, although there are other formation models in
the literature as well \citep{garcia-berro_et_al2012}.

The remainder of this document is structured as follows: In
\S\ref{sec:tides}, I show that the increased moment of inertia of an
evolved star can cause a companion that had been slowly moving outward
due to tidal torques to instead rapidly plunge into the primary.  In
\S\ref{sec:2pops}, I argue that simple binary evolution leads to two
populations of companions to white dwarfs, with a large gap in period
(or orbital radius) between them.  In \S\ref{sec:sdB}, I comment
briefly on the potential sub-Earth-sized planets around KOI 55 found
by \citet{charpinet_et_al2011}.  Finally, in \S\ref{sec:conc}, I
summarize and conclude.

\section{Evolution of Tidal Torques}
\label{sec:tides}

Consider two bodies of masses $M_1$ and $M_2$ each in a circular orbit
around the common center of mass.  The angular momentum of the orbit
is $L_{\rm orb} = \mu a^2 n$, where $\mu \equiv M_1 M_2 / (M_1 + M_2)$
is the reduced mass, $a$ is the orbital semimajor axis, and $n =
2\pi/P$ is the orbital mean motion (where $P$ is the orbital period).
If the moments of inertia of the two bodies are $I_1$ and $I_2$ and
their angular rotation rates are $\Omega_1$ and $\Omega_2$, then their
respective spin angular momenta are $J_1 = I_1 \Omega_1$ and $J_2 =
I_2 \Omega_2$.  The total angular momentum of the system is
\begin{equation}
J = J_1 + J_2 + L_{\rm orb} = I_1 \Omega_1 + I_2 \Omega_2 + \mu a^2 n \, .
\label{eq:Jsys}
\end{equation}
There are no net tidal torques (and therefore the system is tidally
locked) if each member of the binary is spinning at the orbital mean
motion.  In other words, tidal equilibrium fixed points are obtained
when the system's angular momentum is equal
to
\begin{equation}
J = I_1 n_{\rm eq} + I_2 n_{\rm eq} + \mu a_{\rm eq}^2 n_{\rm eq} \, ,
\label{eq:Jeq}
\end{equation}
where $n_{\rm eq}$ and $a_{\rm eq}$ are the equilibrium mean motion
and orbital radius, respectively, for the given system angular
momentum $J$ (in eq.~\ref{eq:Jeq}, the $\Omega$s have been replaced
with $n_{\rm eq}$ because in equilibrium $\Omega=n$).  Since $n =
\sqrt{G (M_1 + M_2) / a^3}$, this may be rewritten as (dropping the
``eq'' subscripts),
\begin{equation}
J = a^{-3/2} \sqrt{G (M_1 + M_2)} \left( I_1 + I_2 + \frac{M_1 M_2}{M_1 + M_2} a^2 \right) \, .
\label{eq:Jeq2}
\end{equation}

Equation~(\ref{eq:Jeq2}) has two striking features.  First, $J$
approaches infinity as $a$ approaches both zero and infinity. Second
(and as a consequence), for any (sufficiently great) total angular
momentum of the system, there are two orbital radii at which the
binary can achieve tidal equilibrium.  This was pointed out by
\citet{darwin1879, darwin1880}, who noted that the Earth-Moon system
could be in tidal equilibrium in a $\sim$5-hour orbit, in addition to
the more commonly recognized $\sim$50-day orbit that we are slowly
approaching.  A perturbative stability analysis shows that the inner
fixed point is unstable --- a ``repeller,'' and the outer one is
stable --- an ``attractor'' \citep{hut1980}.  A system that has an
orbital radius between the two fixed points approaches the outer fixed
point, as the Earth-Moon system is doing.  As pointed out by
\citet{levrard_et_al2009} and others, a system that is inside the
inner fixed point is drawn inexorably by tidal torques toward a merger
(which, in the case of planet-star mergers can produce spectacular
optical, ultraviolet, and X-ray signatures, as described by
\citealt{metzger_et_al2012}).

\begin{figure}[t]
\centerline{\plotone{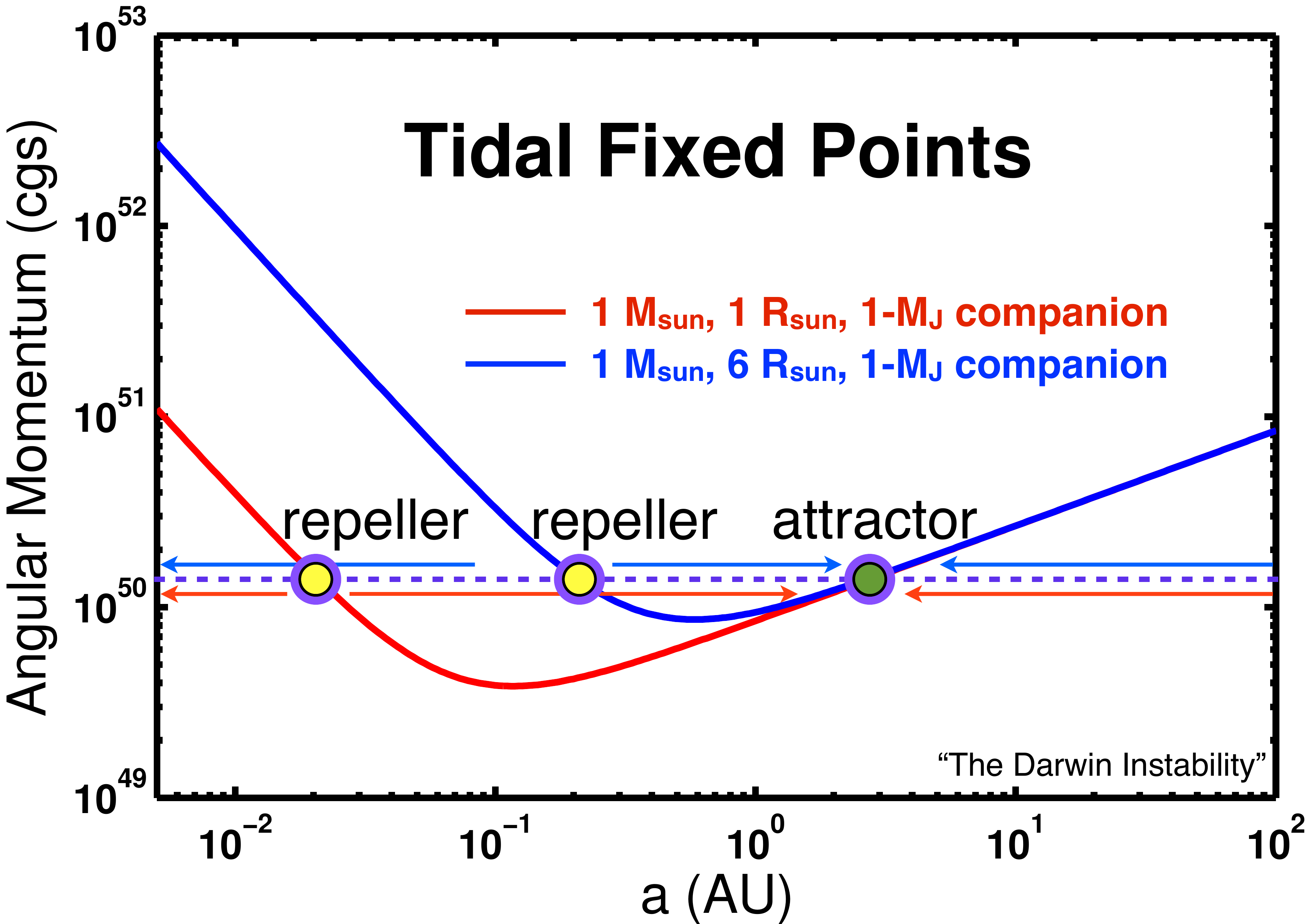}}
\caption{\small Tidal fixed points.  The curves represent orbital
  radius/angular momentum pairs for a 1-$M_\sun$/1-$M_{\rm Jup}$
  binary at which tidal fixed points are possible (i.e., where it is
  possible for both binary components to be spinning with angular
  frequencies equal to the orbital mean motion).  The yellow curve
  represents the locus of fixed points assuming a main-sequence
  solar-mass star as the host; the blue curve represents the locus of
  fixed points assuming a post-main-sequence (6-$R_\sun$) star as the
  host.  The horizontal purple dashed line indicated a particular
  total system angular momentum; note that for this (and for many)
  angular momenta there are two fixed points, a close-in one and a
  more distant one.  The inner one is a repeller and the outer one is
  an attractor.}
\label{fig:angmom1}
\end{figure}

For a system with companion that has a small radius and small mass
relative to the primary, equation~(\ref{eq:Jeq2}) may be simplified as
follows:
\begin{eqnarray}
\label{eq:Jeq_approx1} J & \approx & a^{-3/2} \sqrt{G M_*} \left( I_* + M_c a^2 \right) \\
\label{eq:Jeq_approx2} & = & a^{-3/2} \sqrt{G M_*} \left( \alpha_* M_* R_*^2 + M_c a^2 \right) \, ,
\end{eqnarray}
where the subscripts $*$ and $c$ refer to the primary star and the
companion, respectively, and in equation~(\ref{eq:Jeq_approx2}) the
primary's moment of inertia has been rewritten using $I_* \equiv
\alpha_* M_* R_*^2$ for some value $\alpha_*$.  Although the
$\alpha_*$ value of an evolving star is almost surely a function of
$R_*$ (or alternatively of time), if we take $\alpha_*$ to be constant
then we may examine the approximate behavior of the $J$---$a$ relation
in two limiting cases.  For $M_c a^2 \ll \alpha_* M_* R_*^2$ (i.e.,
near the close-in fixed point),
\begin{equation}
a_{\rm close}[J] \sim G^{1/3} \alpha_*^{2/3} J^{-2/3} M_* R_*^{4/3} \propto R_*^{4/3} \, ;
\label{eq:close}
\end{equation}
whereas for $M_c a^2 \gg \alpha_* M_* R_*^2$ (near the more distant
fixed point),
\begin{equation}
a_{\rm distant}[J] \sim G^{-1} J^{2} M_*^{-1} M_c^{-2} \, ,
\label{eq:farr}
\end{equation}
and is nearly independent of $R_*$.

Figure~\ref{fig:angmom1} illustrates what happens when one member of
the binary (the more massive member) rapidly changes its moment of
inertia --- such as when a star ascends the red giant branch (RGB),
increasing its moment of inertia at constant mass and therefore
constant binary angular momentum.  The yellow and blue curves
represent the set of tidal fixed points for, respectively, a
main-sequence solar-type star, and a post-main-sequence solar-type
star (here, a 6-$R_\sun$ primary with a 1-$M_J$ companion, where $M_J$
is Jupiter's mass).  The inner part of the yellow curve moves outward
in orbital radius, in accordance with equation~(\ref{eq:Jeq_approx1}),
while the outer part remains essentially fixed, as per
equation~(\ref{eq:Jeq_approx2}).

As the repeller point approaches a planet, what happens?  It is
tempting to think that the repeller would push the planet outward, in
the direction of the attractor point.  In truth, though, the outcome
depends on the relative timescales of stellar evolution and tidal
evolution.  The rate of tidal evolution depends on a large power of
the ratio $R_*/a$, where $R_*$ is the primary's radius and $a$ the
companion's orbital radius.  For reasonable assumptions about the
primary's tidal dissipation efficiency, a companion at an orbital
radius greater than a few tenths of an AU has a tidal evolution
timescale that is long compared with the primary's evolution
timescale.  As a result, the primary's evolution and increasing moment
of inertia cause the repeller point to move past the companion.  In
this manner, a system that had been gently evolving toward the outer
(stable) tidal fixed point will suddenly find itself on the inside of
the unstable fixed point, thereupon suffering the ``Darwin
Instability'' and being tidally dragged toward a merger with the star.
This process accelerates as the primary's ascent of the RGB (and
increasing radius) eventually leads to fast tidal evolution.

So long as the tidal evolution timescale is longer than the stellar
evolution timescale, this process is guaranteed to happen for
companions that are close enough to their primaries that the repeller
point would move past the companion.  For realistic assumptions about
tides the orbital evolution will be indeed slower than the stellar
evolution timescale for most companions that are in orbits longer than
a few tens of days.

\begin{figure}[t]
\centerline{\plotone{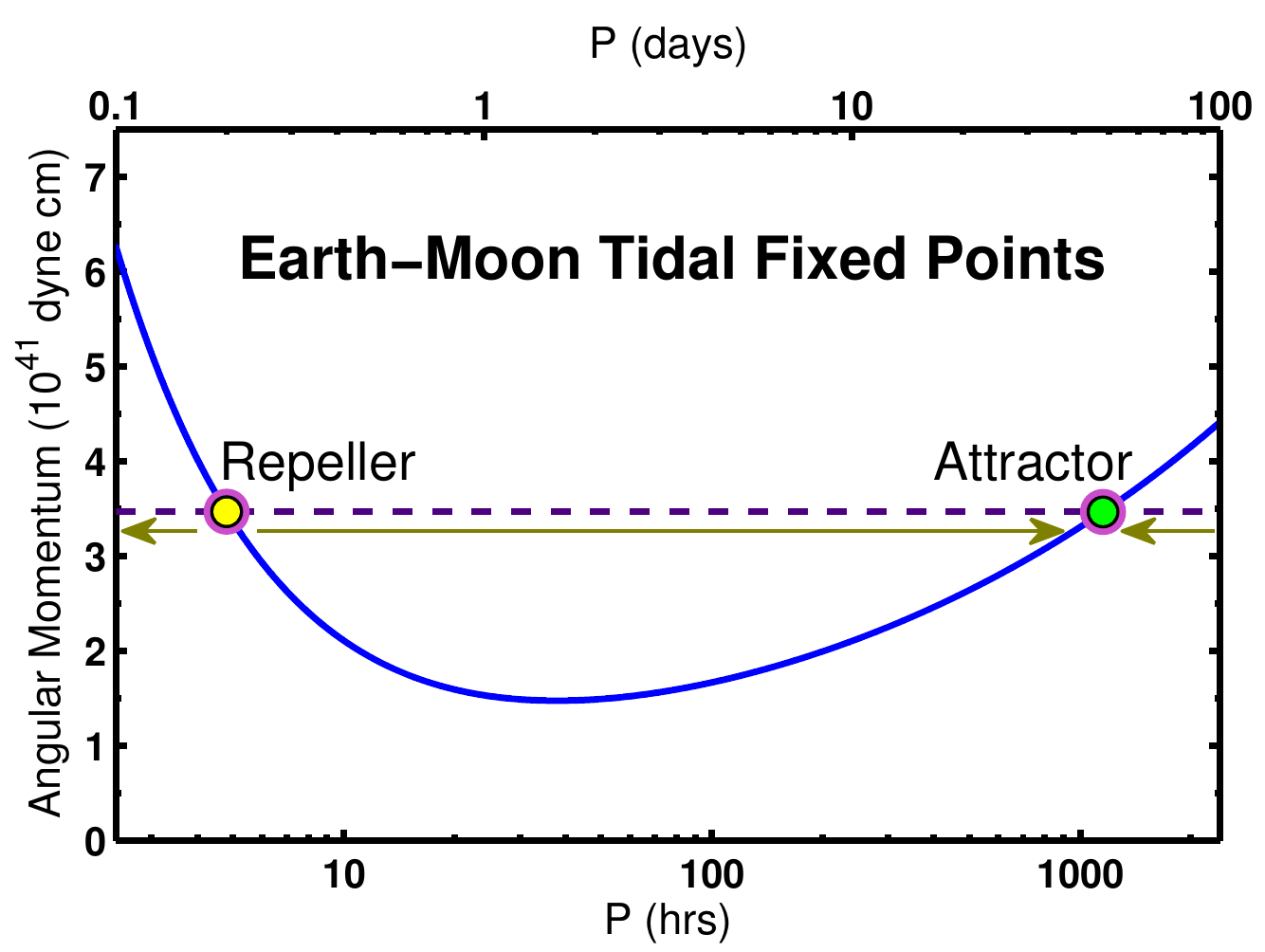}}
\caption{\small Earth-Moon tidal fixed points.  The blue curve
  indicates the locus of orbital period/angular momentum pairs for the
  Earth-Moon system at which tidal fixed points are possible (i.e.,
  where it is possible for both objects to have spin periods equal to
  the orbital period).  The purple dashed curve indicates the system's
  angular momentum.  The Earth-Moon orbit is widening and (slowly)
  approaching the outer (stable) fixed point, the attractor at about
  47 days.}
\label{fig:angmom2}
\end{figure}

For a pedagogical comparison, Fig.~\ref{fig:angmom2} depicts the set
of tidal equilibrium fixed points for the Earth-Moon system.  The
horizontal dashed line indicates Earth-Moon system's angular momentum.
The orbit is evolving under the influence of tides and is slowly
approaching the outer fixed point at an orbital period of about 47
days.

\clearpage

\section{Two Populations}
\label{sec:2pops}

\centerline{\plotone{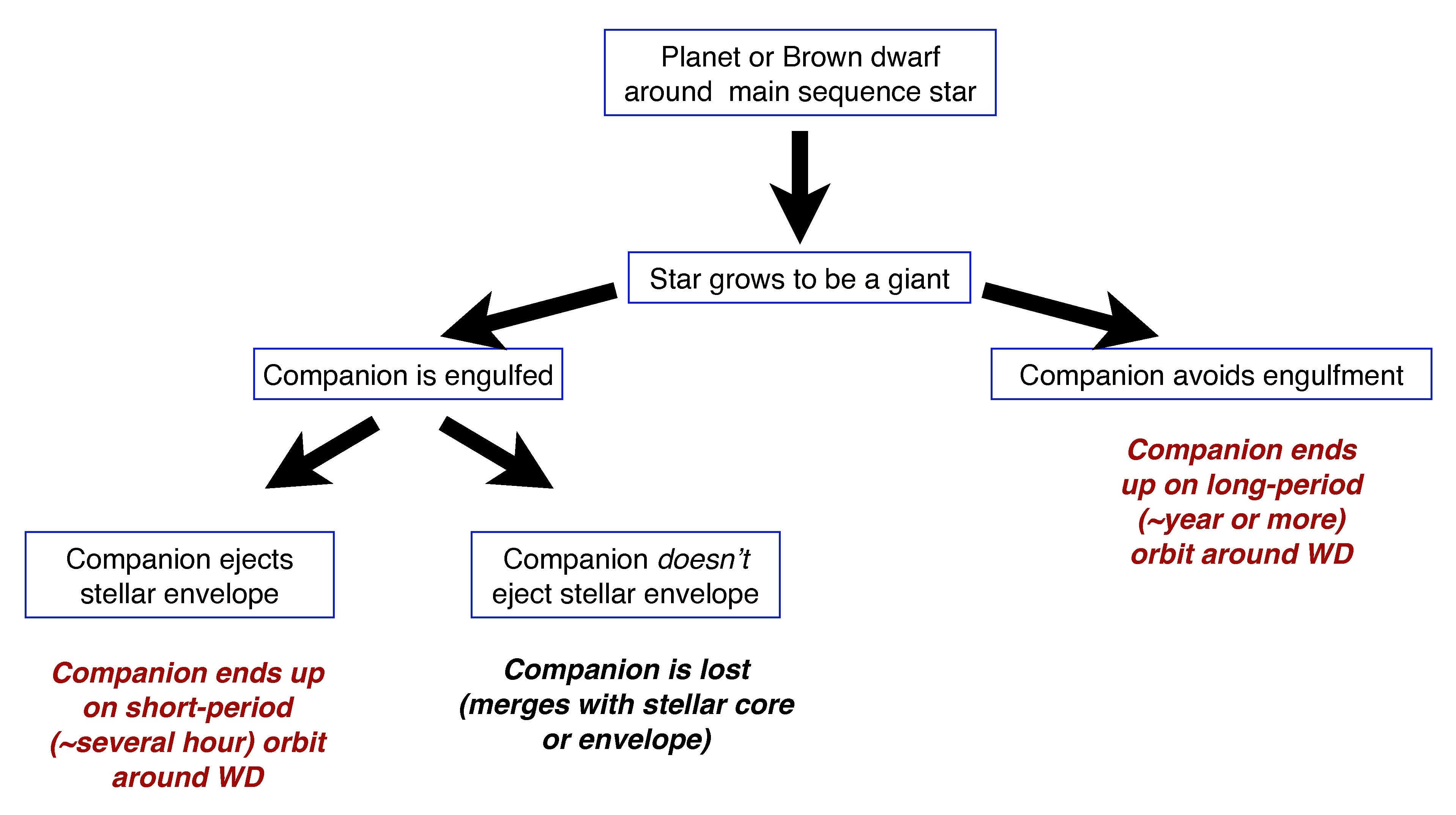}}

The above flowchart summarizes the possible evolutionary paths that a
binary system can follow.  In short, there are three types of
outcomes: a companion can avoid ever being engulfed by the primary; or
a companion can merge with the evolving star and then either survive
the common envelope (CE) phase or be destroyed.

What properties of a binary system affect the eventual outcome?  The
primary's mass loss tends to make make the orbit expand in proportion
to $M_*^{\rm ZAMS}/M_*[t]$, where $M_*^{\rm ZAMS}$ is the zero-age
main-sequence mass of the star, and $M_*[t]$ is its mass at time $t$,
while tidal torques will generally tend to make the orbit shrink (so
long as the orbital timescale is shorter than the evolved star's
rotational period).  Other effects that might be thought to influence
a binary companion's orbit, such as drag from moving through the
enhanced stellar wind, are negligible for planetary-sized (or more
massive) companions \citep{villaver+livio2009}.

Those companions that avoid engulfment must be at least an AU away
from the star (probably actually several-to-20~AU), and those that
survive a CE end up very close ($\lsim$0.01~AU for planetary and
brown-dwarf-mass companions).  As a result, there ought to be a large
gap in orbital separation (or period) in the distribution of binary
companions to white dwarfs.  The approximate boundaries of the gap may
be estimated as described below:

If a companion is to avoid ever being swallowed, it must be far enough
from the primary that tidal torques cannot sap it of enough orbital
energy to make it plunge.  Several recent works have investigated the
set of initial planetary orbits that lead to mergers
\citep{carlberg_et_al2009, villaver+livio2009, nordhaus_et_al2010,
  mustill+villaver2012, nordhaus+spiegel2013} and found that if a
several-Jupiter-mass object begins at least a few times the maximum
stellar radius achieved during the RGB or asymptotic giant branch
(AGB) phase, it will avoid being tidally engulfed (where ``a few''
means $\sim$3---6, for a fiducial stellar tides prescription, as
calibrated by \citealt{verbunt+phinney95}).  Since the maximum stellar
radius for a 1---3-$M_\sun$ star is in the range of one to several AU,
this suggests that the outer boundary of the gap ranges from a few to
$\sim$20 AU, depending on the progenitor star's main-sequence mass.

\begin{figure}[t!]
\centerline{\plotone{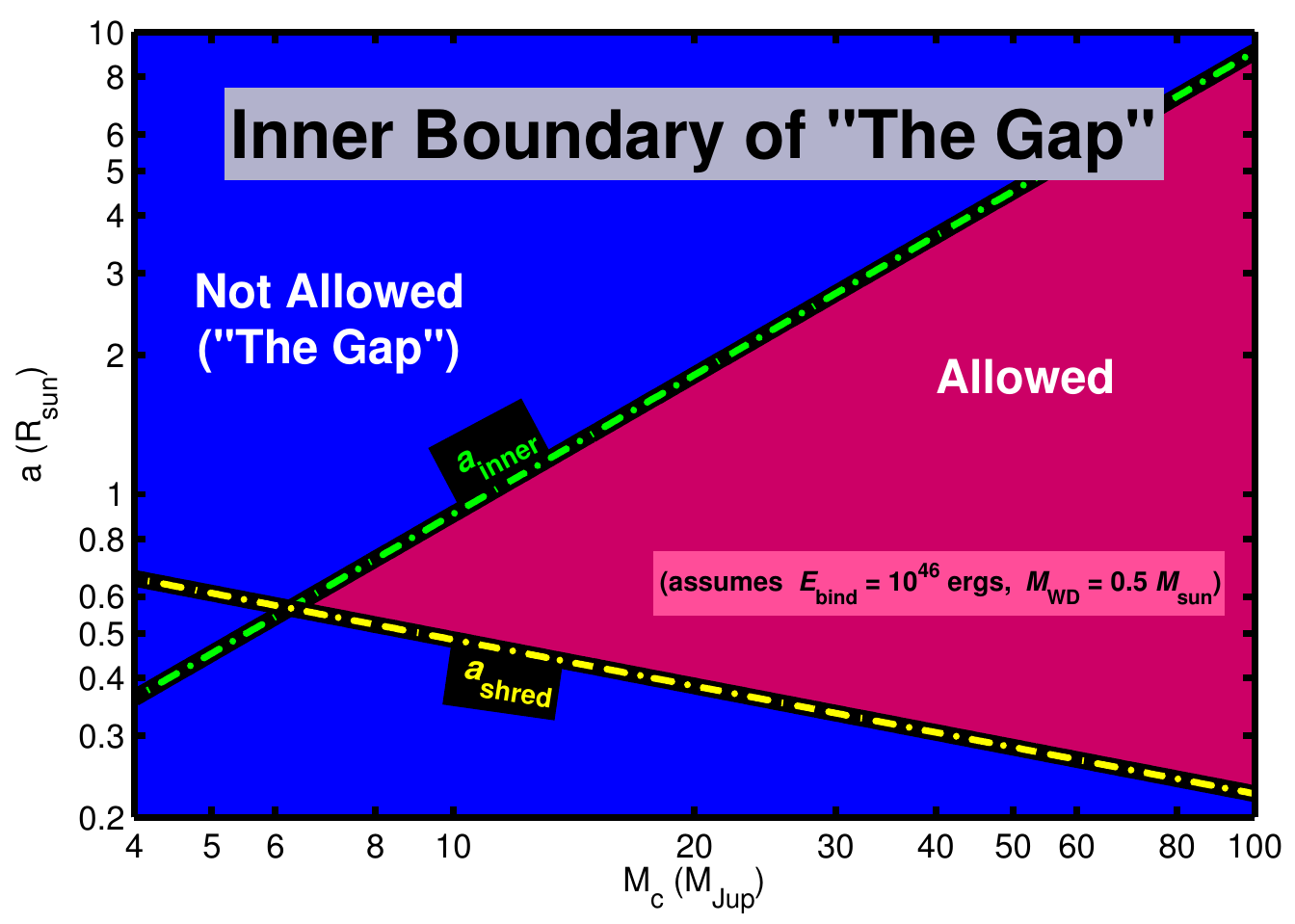}}
\caption{\small Interior to the period/orbital-radius gap for binary
  companions to white dwarfs.  The blue region is inaccessible to
  binary companions.  The magenta region, consisting of companions
  more massive than $\sim$6~$M_J$ and inside a few $R_\sun$ in orbital
  radius, is energetically allowed for binary companions and yet is
  far enough from the white dwarf that a 1-$R_J$ companion will not be
  tidally shredded.  This figure assumes an envelope binding energy of
  $10^{46}$~ergs and a white dwarf mass of 0.5~$M_\sun$.  Higher
  envelope binding energies result in companions that are closer to
  the white dwarf (and in a greater minimum companion mass).  The
  green dashed-dotted line indicates $a_{\rm inner}$
  (equation~\ref{eq:ainner2}), below which a companion can dissipate
  enough orbital energy to unbind a $10^{46}$-erg envelope, and the
  yellow dashed-dotted line indicates $a_{\rm shred}$
  (equation~\ref{eq:ashred2}), above which a 1-$R_J$ companion avoids
  tidal disruption.}
\label{fig:agap}
\end{figure}

If a companion is to survive a common envelope \citep{soker_et_al1984,
  dewi+tauris2000, passy_et_al2012a, passy_et_al2012b}, it must be
massive enough that the energy it injects into the stellar envelope is
sufficient to unbind the star before the companion is tidally
disrupted.  If the degenerate core's mass is $M_{\rm WD}$, then the
orbital energy lost as the companion sinks through the CE to an
orbital radius of $a$ is $\Delta E_{\rm orb} \sim -GM_{\rm WD}M_c/2a$.
Therefore, if the stellar envelope's binding energy is $E_{\rm bind}$,
the radius of the inner edge of the gap may be approximated as
\newpage
\begin{eqnarray}
\label{eq:ainner1} a_{\rm inner} & \sim & \frac{G M_{\rm WD} M_c}{2 E_{\rm bind}} \\
\label{eq:ainner2} & \sim & 0.9 R_\odot \left( \frac{M_c}{10 M_{\rm Jup}} \right) \left( \frac{M_{\rm WD}}{0.5 M_\odot} \right) \left( \frac{E_{\rm bind}}{10^{46} \rm~erg} \right)^{-1} \, .
\end{eqnarray}
Equation~(\ref{eq:ainner2}) assumes that all the lost orbital energy
goes into unbinding the stellar envelope.  If only a fraction
$\alpha_{\rm CE}$ of this energy actually participates in unbinding
the envelope (for further details on CE alpha, see
\citealt{soker2012}), then the actual maximum orbital radius at the
inner edge of the gap is reduced by a factor of $\alpha_{\rm CE}$.
Similarly, if, for some system, $\alpha_{\rm CE}$ were greater than
unity, the maximum orbital radius at the inner edge of the gap could
be greater than that indicated in equation~(\ref{eq:ainner2}), but I am
aware of no suggestion that $\alpha_{\rm CE}$ could be dramatically
larger than 1 for a companion that survives the CE.\footnote{If a
  hydrogen-rich companion, such as a Jupiter, falls deep enough into
  an AGB star's interior, this could deposit fresh hydrogen in a
  helium-burning layer, which might help to donate significantly more
  energy to the envelope than simply the lost orbital energy, but at
  the cost of the companion's survival, as noted by
  \citet{nordhaus_et_al2011}.}  The orbital period at the inner edge
of the gap is
\begin{eqnarray}
\nonumber P_{\rm inner} & \sim & \left( \frac{4 \pi^2 a_{\rm inner}^3}{ G M_{\rm WD}} \right)^{1/2} \\
\label{eq:Pinner} & \sim & {3.4 \rm~hrs} \times \left( \frac{M_c}{10 M_{\rm Jup}} \right)^{3/2} \left( \frac{M_{\rm WD}}{0.5 M_\odot} \right) \left( \frac{E_{\rm bind}}{10^{46} \rm~erg} \right)^{-3/2} \, .
\end{eqnarray}

Figure~\ref{fig:agap} illustrates the region of parameter space that
is allowed at the inner edge of the gap, as a function of companion
mass and orbital radius.  Not only is there a maximum orbital radius
for the inner edge of the gap, but there is a minimum radius too,
defined by the tidal shredding radius
\begin{eqnarray}
\label{eq:ashred1} a_{\rm shred} & \sim & R_c \left(\frac{2 M_{\rm WD}}{M_c}\right)^{1/3} \\
\label{eq:ashred2} & \sim & 0.5 R_\sun \times \left( \frac{R_c}{R_J} \right) \left( \frac{M_{\rm WD}}{0.5 M_\sun} \right)^{1/3} \left( \frac{M_c}{10 M_J} \right)^{-1/3}  \, ,
\end{eqnarray}
where $R_c$ is the companion's radius and $R_J$ is Jupiter's radius.
The maximum and minimum orbital radii ($a_{\rm inner}$ and $a_{\rm
  shred}$) are equal for a companion mass of
\begin{eqnarray}
\label{eq:Mcmin1} {M_c}^{\rm min} & \sim &  \left( \frac{16 R_c^3 E_{\rm bind}^3}{G^3 M_{\rm WD}^2} \right)^{1/4} \\
\label{eq:Mcmin2} & \sim & 6 M_J \times \left( \frac{R_c}{R_J} \right)^{3/4} \left( \frac{E_{\rm bind}}{10^{46} \rm~erg} \right)^{3/4} \left( \frac{M_{\rm WD}}{M_\sun} \right)^{-1/2} \, .
\end{eqnarray}
Companions less massive than this ought to be shredded and
incorporated into the stellar envelope before they lose enough orbital
energy to unbind the stellar envelope, and consequently it would be a
mystery if they were to survive the CE phase (although, were low-mass
planets to form after the formation of the white dwarf, they might
still be found inside the gap region).  This conclusion is consistent
with the results of \citet{nelemans+tauris1998}, who found that
planet-mass objects are unlikely to survive CE evolution and end up in
a close orbit around a WD.

\section{sdB Planets}
\label{sec:sdB}

\citet{charpinet_et_al2011} recently announced the discovery of a
puzzling object in the Kepler data.  KOI-55 is a subdwarf-B (sdB) star
of radius $\sim$0.2~$R_\sun$ whose lightcurve shows regular variations
with two distinct periods ($\sim$8.2 hours and $\sim$5.8) in a 10:7
ratio.  \citeauthor{charpinet_et_al2011} argue that these variations
are inconsistent with any intrinsic pulsational periods in a star of
this type and that the most plausible explanation is the presence of
two sub-Earth-radius (nontransiting) planets whose daysides
periodically come into view.

It is difficult to see how these putative objects could have arrived
at their orbits through ordinary CE evolution.  For stability reasons,
their masses must be significantly less than Jupiter's, as noted by
\citeauthor{charpinet_et_al2011} and as can easily be verified with an
N-body integrator such as, e.g., {\tt REBOUND} \citep{rein+liu2012}.
For instance, if they are as massive as Jupiter, one of them would
probably be ejected from the system in less than a year.  Since they
must be significantly less massive than Jupiter, the objects that are
currently in orbit around the star could not have deposited enough
orbital energy in the stellar envelope to unbind it.  The same
stability argument suggests that it would require extreme fine tuning
for two massive ($\gtrsim$10~$M_J$) bodies to move through a common
envelope together, in resonance, and evaporate down to
$\sim$Earth-sized cores at their present locations.
\citet{bear+soker2012} argue that these objects might be the result of
a single massive companion that was tidally shredded during a CE
phase.  In this scenario, the core was shredded, too, and two chunks
of the core were flung from the shredding radius to their current
orbits.  Another conceivable scenario, if these lightcurve variations
correspond to actual companions, is that both objects migrated to
their current locations after the sdB star formed from a previous CE
companion that was destroyed during the CE phase.\footnote{This
  migration process might have occurred via, e.g., Kozai interactions
  followed by tidal circularization \citep{kozai1962,
    fabrycky+tremaine2007, katz_et_al2011, socrates_et_al2012,
    naoz_et_al2012, shappee+thompson2012}.}  This scenario would
require an outer perturber, and would not be a result of simple binary
CE evolution.  The planet hypothesis for this puzzling lightcurve
might be true, but it is not obvious how the planets could have
reached their present orbits.

\section{Conclusion}
\label{sec:conc}

Post-main-sequence stellar evolution results in dramatic changes in
the stellar radius and, therefore, in the orbits of companions to the
stars.  Companions that are too close can either be directly swallowed
by the expanding star or tidally dragged into merging with the star.
The ensuing common envelope poses severe risks to the survival of the
companion; massive enough companions can eventually transfer enough
orbital energy to unbind the star, resulting in a tight post-CE orbit,
while less massive companions continue spiraling inwards until they
are tidally shredded and merge with their host stars.  Companions that
are on distant enough orbits to avoid ever merging with their host
stars move outward due to mass loss from the primary.

Nearly all objects that are massive enough to survive a CE have masses
in excess of the deuterium-burning limit that is sometimes used to
dilineate between planets and brown dwarfs \citep{spiegel_et_al2011a}.
For an object that is less than $\sim$6 times the mass of Jupiter to
end up inside 1~AU from a WD requires something more complex than
simple binary/CE evolution.  In particular, forming the potentially
habitable circum-white-dwarf terrsstrial objects considered by
\citet{agol2011} and \citet{fossati_et_al2012} might require exotic
circumstances involving multiple bodies, and the evolutionary paths to
produce such worlds might be problematic for their subsequent
habitability, as pointed out by \citet{nordhaus+spiegel2013}.

\vspace{1in}

{\sc Acknowledgments}

This work was in part inspired by the ``Planets Around Stellar
Remnants'' conference in Arecibo, Puerto Rico, January 23-27, 2012.  I
thank Jason Nordhaus, John Johnson, Ruobing Dong, Scott Gaudi, Jeremy
Goodman, and Piet Hut for illuminating conversations, and Alex
Wolszczan for organizing the conference.  I gratefully acknowledge
support from NSF grant AST-0807444 and the Keck Fellowship, and from
the Friends of the Institute.

\bibliography{biblio.bib}

\end{document}